\documentstyle[sprocl,epsfig]{article}

\def\Journal#1#2#3#4{{#1} {\bf #2}, #3 (#4)}

\def\NPA{{\em Nucl. Phys.} A}

\def\PLB{{\em Phys. Lett.}  B}
\def\PRL{\em Phys. Rev. Lett.}
\def\PRC{{\em Phys. Rev.} C}
\def\PRD{{\em Phys. Rev.} D}

\def\be{\begin{equation}}
\def\ee{\end{equation}}
\def\bea{\begin{eqnarray}}
\def\eea{\end{eqnarray}}
\newcommand{\case}[2]{\mbox{\footnotesize $\displaystyle \frac{#1}{#2}$}}
\newcommand{\lsim}{\mathrel{\rlap{\lower4pt\hbox{\hskip0pt$\sim$}}
\raise2pt\hbox{$<$}}}
\newcommand{\gsim}{\mathrel{\rlap{\lower4pt\hbox{\hskip0pt$\sim$}}
\raise2pt\hbox{$>$}}}

\begin{document}

\title{QCD BOUND STATES AND THEIR RESPONSE TO EXTREMES OF TEMPERATURE AND
DENSITY} 

\author{P. MARIS AND C. D. ROBERTS}

\address{Physics Division, Bldg. 203, Argonne National Laboratory\\
Argonne IL 60439-4843, USA}
% \\
% E-mail: maris@theory.phy.anl.gov, cdroberts@anl.gov}


\maketitle\abstracts{We describe the application of Dyson-Schwinger equations
to the calculation of hadron observables.  The studies at zero temperature
$(T)$ and quark chemical potential $(\mu)$ provide a springboard for the
extension to finite-$(T,\mu)$.  Our exemplars highlight that much of hadronic
physics can be understood as simply a manifestation of the nonperturbative,
momentum-dependent dressing of the elementary Schwinger functions in QCD.}

\section{DSE Essentials}
The Dyson-Schwinger equations (DSEs) provide a Poincar\'e invariant,
continuum approach to solving quantum field theories.  They are an infinite
tower of coupled integral equations, with the equation for a particular
$n$-point function involving at least one $m>n$-point function.  A tractable
problem is only obtained if one truncates the system, and historically this
has provided an impediment to the application of DSEs: {\it a priori} it can
be difficult to judge whether a particular truncation scheme will yield
qualitatively or quantitatively reliable results for the quantity sought.  As
integral equations, the analysis of observables is a numerical problem and
hence a critical evaluation of truncation schemes often requires access to
high-speed computers.\footnote{The human and computational resources required
are still modest compared with those consumed in contemporary numerical
simulations of lattice-QCD.}  With such tools now commonplace, this
evaluation can be pursued fruitfully.

The development of efficacious truncation schemes in not a purely numerical
task, and neither is it always obviously systematic.  For some, this last
point diminishes the appeal of the approach.  However, with growing community
involvement and interest, the qualitatively robust results and intuitive
understanding that the DSEs can provide is becoming clear.  Indeed, someone
familiar with the application of DSEs in the late-70s and early-80s might be
surprised with the progress that has been made.  It is now
clear$\,$\cite{bender96} that truncations which preserve the global
symmetries of a theory; for example, chiral symmetry in QCD, are relatively
easy to define and implement and, while it is more difficult to preserve
local gauge symmetries, much progress has been made with Abelian
theories$\,$\cite{ayse97} and more is being learnt about non-Abelian ones.

The simplest truncation scheme for the DSEs is the weak-coupling expansion.
It shows that the DSEs {\it contain} perturbation theory; i.e, for any given
theory the weak-coupling expansion generates all the diagrams of perturbation
theory.  However, the most important feature of the DSEs is the antithesis of
this weak-coupling expansion: the DSEs are intrinsically nonperturbative and
their solution contains information that is {\it not} present in perturbation
theory.  They are ideal for the study of dynamical chiral symmetry breaking
(DCSB) and confinement in QCD, and of hadronic bound state structure and
properties.  In this application they provide a means of elucidating
identifiable signatures of the quark-gluon substructure of hadrons.

Their intrinsically nonperturbative nature also makes them well suited to
studying QCD at finite-$(T,\mu)$, where the characteristics of the phase
transition to a quark-gluon plasma are a primary subject.  The order of the
transition, the critical exponents, and the response of bound states to
changes in these intensive variables: all must be elucidated.  The latter
because there lies the signals that will identify the formation of the plasma
and hence guide the current and future experimental searches.

\subsection{Gluon Propagator}\label{subsec:B1}
In Landau gauge the two-point, dressed-gluon Schwinger function, or
dressed-gluon propagator, has the form
\begin{eqnarray}
g^2 D_{\mu\nu}(k)& = &
\left(\delta_{\mu\nu} - \frac{k_\mu k_\nu}{k^2}\right)
        \frac{{\cal G}(k^2)}{k^2}\,,\; 
        {\cal G}(k^2):= \frac{g^2}{[1+\Pi(k^2)]} \,,
\end{eqnarray}
where $\Pi(k^2)$ is the vacuum polarisation, which contains all the dynamical
information about gluon propagation.  Studies of the gluon DSE have been
reported by many authors$\,$\cite{rw94} with the conclusion that, if the
ghost-loop is unimportant, then the charge-antiscreening 3-gluon vertex
dominates and, relative to the free gauge boson propagator, the dressed gluon
propagator is significantly enhanced in the vicinity of $k^2=0$.\footnote{
The possibility that ${\cal G}(k^2)$ is finite or vanishes at $k^2=0$ is
canvassed in these proceedings.  In the absence of particle-like
singularities in the quark-gluon vertex such behaviour is very difficult to
reconcile with the observable phenomena of QCD.$\,$\protect\cite{hawes} A
particle-like singularity is one of the form $(P^2)^{-\alpha}$, $\alpha \in
(0,1]$. In this case one can write a spectral decomposition for the vertex in
which the spectral densities are non-negative.  This is impossible if
$\alpha>1$. $\alpha=1$ is the ideal case of an isolated, $\delta$-function
singularity in the spectral densities and hence an isolated, free-particle
pole.  $\alpha \in (0,1)$ corresponds to an accumulation, at the particle
pole, of branch points associated with multiparticle production.}
The enhancement persists to $k^2 \sim 1$-$2\,$GeV$^2$, where a perturbative
analysis becomes quantitatively reliable.  In the neighbourhood of $k^2=0$
the enhancement can be represented$\,$\cite{bp89} as a regularisation of
$1/k^4$ as a distribution.  A dressed-gluon propagator of this type generates
confinement\footnote{
One aspect of confinement is the absence of quark and gluon production
thresholds in colour-singlet-to-singlet ${\cal S}$-matrix amplitudes.  This
is ensured if the dressed-quark and -gluon propagators do not have a Lehmann
representation.}
and DCSB {\it without} fine-tuning.
\subsection{Quark Propagator}\label{subsec:B2}
In a covariant gauge the two-point, dressed-quark Schwinger function, or
dressed-quark propagator, can be written in a number of equivalent forms
\begin{eqnarray}
\label{Sp}
S(p) & := & \frac{1}{i\gamma\cdot p + \Sigma(p)} \\ 
& := & \frac{1}{i\gamma\cdot p\,A(p^2) + B(p^2)} 
        \equiv -i\gamma\cdot p \,\sigma_V(p^2) + \sigma_S(p^2) \,.
\end{eqnarray}
$\Sigma(p)$ is the dressed-quark self energy, which satisfies
\begin{equation}
\label{gendse}
\Sigma(p)  =  ( Z_2 -1)\, i\gamma\cdot p + Z_4\,m_{\rm bm}
+\, Z_1\, \int^\Lambda_q \,
g^2 D_{\mu\nu}(p-q) \frac{\lambda^a}{2}\gamma_\mu S(q)
\Gamma^a_\nu(q,p) \,,
\end{equation}
where $\Gamma^a_\nu(q;p)$ is the renormalised dressed-quark-gluon vertex,
$m_{\rm bm}$ is the Lagrangian current-quark bare mass and $\int^\Lambda_q :=
\int^\Lambda d^4 q/(2\pi)^4$ represents mnemonically a {\em
translationally-invariant} regularisation of the integral, with $\Lambda$ the
regularisation mass-scale.  The quark-gluon-vertex and quark wave function
renormalisation constants, $Z_1(\mu^2,\Lambda^2)$ and $Z_2(\mu^2,\Lambda^2)$,
depend on the renormalisation point, $\mu$, and the regularisation
mass-scale, as does the mass renormalisation constant $Z_m(\mu^2,\Lambda^2)
:= Z_2(\mu^2,\Lambda^2)^{-1} Z_4(\mu^2,\Lambda^2)$.

\begin{figure}[t]
\centering{\ 
\epsfig{figure=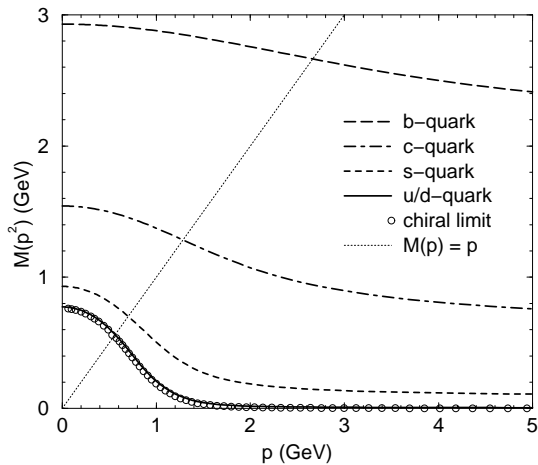,height=7.0cm,rheight=6.3cm}}%\vspace*{0.3\baselineskip}
\caption{Dressed-quark mass-function obtained in solving the quark DSE.
\label{plotMpp}}
\end{figure}

The quark mass-function is $M(p^2):= B(p^2)/A(p^2)$
% \begin{equation}
% \label{Mpp}
% M(p^2) := \frac{B(p^2)}{A(p^2)}
% \end{equation}
and, as illustrated in Fig.~\ref{plotMpp},$\,$\cite{mr97}
solving the quark DSE using an infrared-enhanced dressed-gluon propagator and
a dressed-quark-gluon vertex, $\Gamma_\mu(p,q)$, that does not exhibit
particle-like singularities at $(p-q)^2=0$, one obtains a quark mass-function
that mirrors the infrared enhancement of the dressed-gluon propagator.  These
results were obtained using the one-loop formula for the running mass, with
current-quark masses corresponding to
\begin{equation}
\label{monegev}
\begin{array}{llll}
m_{u/d}^{1\,{\rm GeV}} &
m_s^{1\,{\rm GeV}}     & 
m_c^{1\,{\rm GeV}}     & 
m_b^{1\,{\rm GeV}}     \\
 6.6\, {\rm MeV} & 
 140\,{\rm MeV}  & 
 1.0\,{\rm GeV}  & 
 3.4\,{\rm GeV}\,.
\end{array}
\end{equation}

The quark DSE was also solved in the chiral limit, which in QCD is obtained
by setting the Lagrangian current-quark bare mass to zero.\cite{mr97} One
observes immediately that the mass-function is nonzero even in this case.
That {\it is} DCSB: a momentum-dependent quark mass, generated dynamically,
in the absence of any term in the action that breaks chiral symmetry
explicitly.  This entails a nonzero value for the quark condensate in the
chiral limit.  That $M(p^2)\neq 0$ in the chiral limit is independent of the
details of the infrared-enhanced dressed-gluon propagator.

Figure~\ref{plotMpp} illustrates that for light quarks ($u$, $d$ and $s$)
there are two distinct domains: perturbative and nonperturbative.  In the
perturbative domain the magnitude of $M(p^2)$ is governed by the the
current-quark mass. For $p^2< 1\,$GeV$^2$ the mass-function rises sharply.
This is the nonperturbative domain where the magnitude of $M(p^2)$ is
determined by the DCSB mechanism; i.e., the enhancement in the dressed-gluon
propagator.  This emphasises that DCSB is more than just a nonzero value of
the quark condensate in the chiral limit!  

The solution of $p^2=M^2(p^2)$ defines a Euclidean constituent-quark mass,
$M^E$.  For a given quark flavour, the ratio $M^E_f/m_f^\mu$ is a single,
quantitative measure of the importance of the DCSB mechanism in modifying the
quark's propagation characteristics.  As illustrated in Eq.~(\ref{Mmratio}),
\begin{equation}
\label{Mmratio}
\begin{array}{l|c|c|c|c|c}
\mbox{\sf flavour} 
        &   u/d  &   s   &  c  &  b  &  t \\\hline
 \frac{M^E}{m_{\mu\sim 20\,{\rm GeV}}}
       &  150   &    10      &  2.3 &  1.4 & \to 1
\end{array}
\end{equation}
this ratio provides for a natural classification of quarks as either light or
heavy.  For light-quarks the ratio is characteristically $10$-$100$ while for
heavy-quarks it is only $1$-$2$.  The values of this ratio signal the
existence of a characteristic DCSB mass-scale: $M_\chi$. At $p^2>0$ the
propagation characteristics of a flavour with $m_f^\mu< M_\chi$ are altered
significantly by the DCSB mechanism, while for flavours with $m_f^\mu\gg
M_\chi$ it is irrelevant, and explicit chiral symmetry breaking dominates.
It is apparent from the figure that $M_\chi \sim 0.2\,$GeV$\,\sim
\Lambda_{\rm QCD}$.  This forms the basis for simplifications in the study of
heavy-meson observables.\cite{misha}
\subsection{Hadrons: Bound States}\label{subsec:B3}
The properties of hadrons can be understood in terms of their substructure by
studying covariant bound state equations: the Bethe-Salpeter equation (BSE)
for mesons and the covariant Fadde'ev equation for baryons.  The mesons have
been studied most extensively and their internal structure is described by a
Bethe-Salpeter amplitude obtained as a solution of
\begin{eqnarray}
\label{genbse}
\left[\Gamma_H(k;P)\right]_{tu} &= & 
\int^\Lambda_q  \,
[\chi_H(q;P)]_{sr} \,K^{rs}_{tu}(q,k;P)\,,
\end{eqnarray}
where $\chi_H(q;P) := {\cal S}(q_+) \Gamma_H(q;P) {\cal S}(q_-)$; ${\cal
S}(q) = {\rm diag}(S_u(q),S_d(q),S_s(q), \ldots)$; $q_+=q + \eta_P\, P$,
$q_-=q - (1-\eta_P)\, P$, with $P$ the total momentum of the bound state and
observables independent of $\eta_P$; and $r$,\ldots,$u$ represent colour-,
Dirac- and flavour-matrix indices.  The amplitude for a pseudoscalar bound
state has the form
\begin{eqnarray}
\label{genpibsa}
\Gamma_H(k;P) & = &  T^H \gamma_5 \left[ i E_H(k;P) + 
\gamma\cdot P F_H(k;P) \rule{0mm}{5mm}\right. \\
\nonumber & & 
\left. \rule{0mm}{5mm}+ \gamma\cdot k \,k \cdot P\, G_H(k;P) 
+ \sigma_{\mu\nu}\,k_\mu P_\nu \,H_H(k;P) 
\right]\,,
\end{eqnarray}
where $T^H$ is a flavour matrix that determines the mesonic channel under
consideration; e.g., $T^{K^+}:= (1/2)\left(\lambda^4 + i \lambda^5\right)$,
with $\{\lambda^j,j=1\ldots 8\}$ the Gell-Mann matrices.  In
Eq.~(\ref{genbse}), $K$ is the renormalised, fully-amputated, quark-antiquark
scattering kernel and important in the successful application of DSEs is that
it has a systematic skeleton expansion in terms of the elementary,
dressed-particle Schwinger functions; e.g., the dressed-quark and -gluon
propagators.
\begin{figure}[t]
  \centering{\ \hspace*{-2.5cm}\epsfig{figure=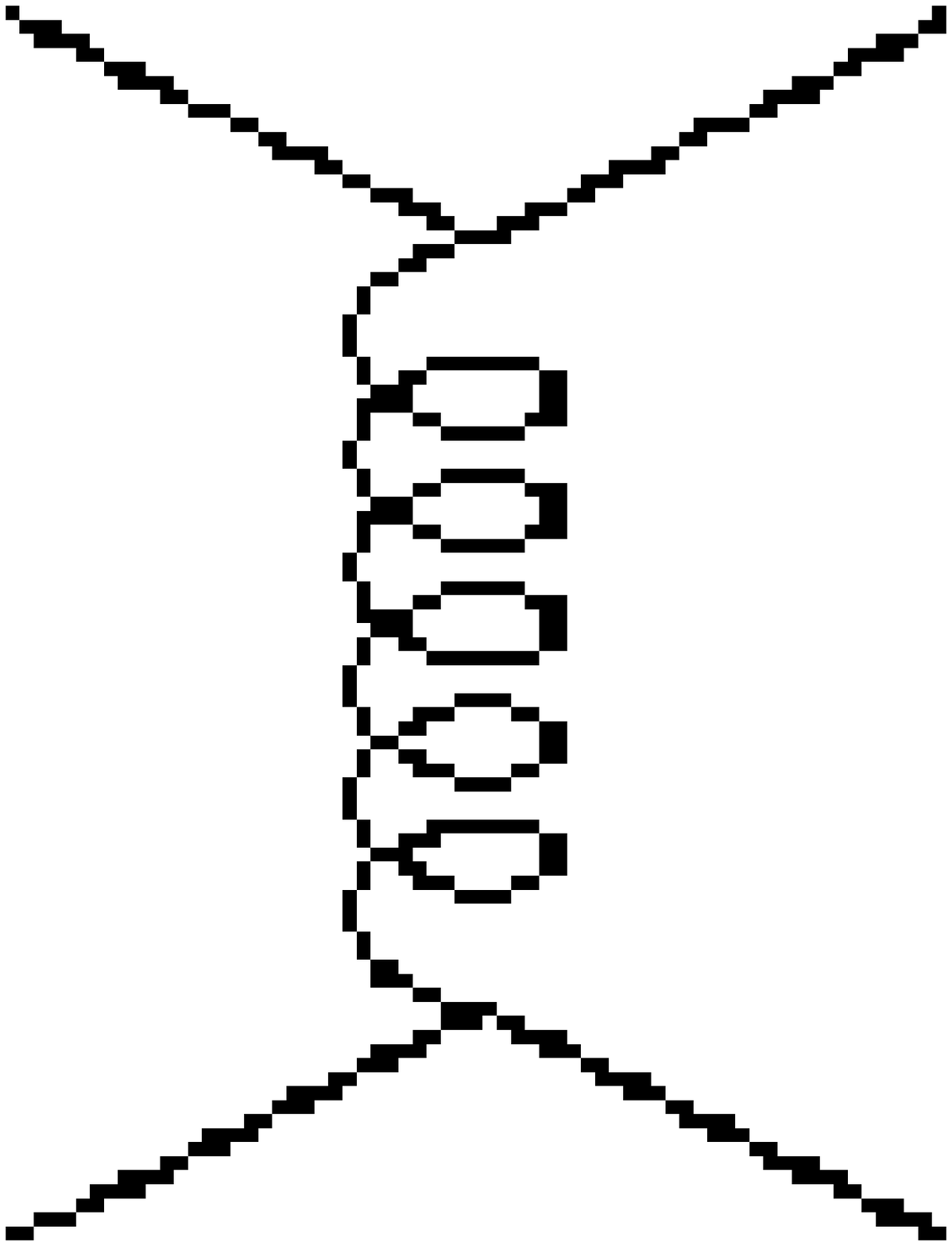,height=2.0cm} 
        \vspace*{-15mm} 

        \hspace*{15mm} $\longleftarrow$ { (1) -- Ladder}\vspace*{4mm} 

        \hspace*{55mm} { (2) -- Beyond Ladder}

        \hspace*{16mm}$\swarrow$

        \hspace*{0.5cm}\epsfig{figure=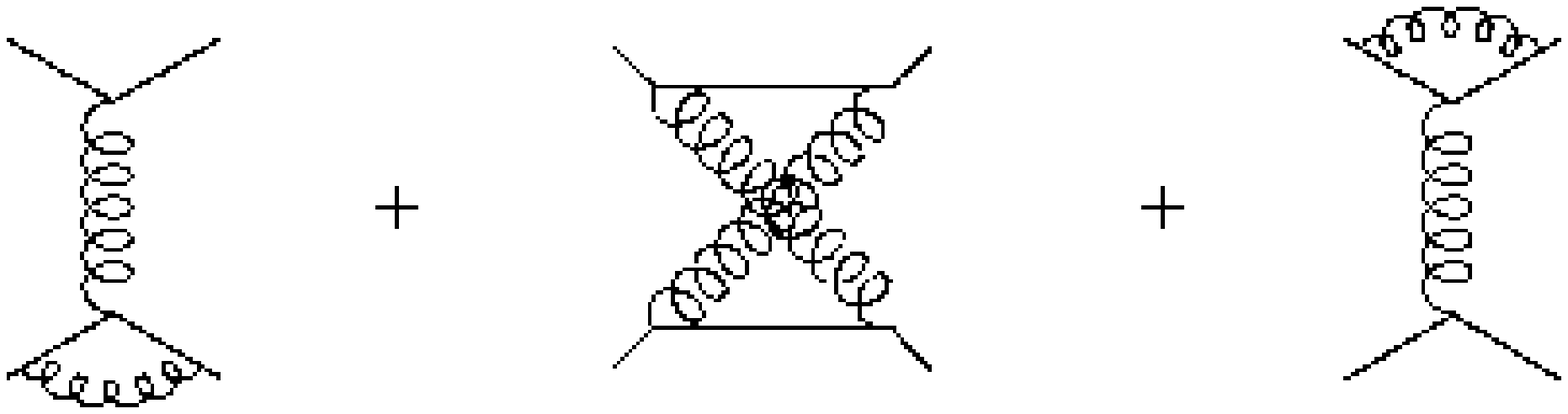,height=2.0cm} }
\caption{First two orders in a systematic expansion of the quark-antiquark
scattering kernel.  In this expansion, the propagators are dressed but the
vertices are bare.
\label{skeleton}}
\end{figure}
This particular expansion,\cite{bender96} with its analogue for the kernel in
the quark DSE, provides a means of constructing a kernel that, order-by-order
in the number of vertices, ensures the preservation of vector and
axial-vector Ward-Takahashi identities.

In any study of meson properties, one chooses a truncation for $K$.  The BSE
is then fully specified and straightforward to solve, yielding the bound
state mass and amplitude.  The ``ladder'' truncation of $K$ combined with the
``rainbow'' truncation of the quark DSE [$\Gamma_\mu \to \gamma_\mu$ in
Eq.~(\ref{gendse})] is the simplest and most often used.  The expansion of
Fig.~\ref{skeleton} provides the explanation$\,$\cite{bender96} for why this
Ward-Takahashi identity preserving truncation is accurate for
flavour-nonsinglet pseudoscalar and vector mesons: there are cancellations
between the higher-order diagrams.  It also shows why it provides a poor
approximation in the study of scalar mesons, where the higher-order terms do
not cancel, and for flavour-singlet mesons, where it omits timelike gluon
exchange diagrams.
\section{A Mass Formula}\label{sec:C}
The dressed-axial-vector vertex satisfies a DSE whose kernel is $K$, and
because of the systematic expansion described in Sec.~\ref{subsec:B3} it
follows$\,$\cite{mr97} that the axial-vector Ward-Takahashi identity
(AV-WTI):
\begin{eqnarray}
\label{avwti}
\lefteqn{-i P_\mu \Gamma_{5\mu}^H(k;P)  = }\\
&& \nonumber
{\cal S}^{-1}(k_+)\gamma_5\frac{T^H}{2}
+  \gamma_5\frac{T^H}{2} {\cal S}^{-1}(k_-) 
- M_{(\mu)}\,\Gamma_5^H(k;P) - \Gamma_5^H(k;P)\,M_{(\mu)} \,,
\end{eqnarray}
[$M_{(\mu)}= {\rm diag}(m_u^\mu,m_d^\mu,m_s^\mu,\ldots)$ is the current-quark
mass matrix] is satisfied in any thoughtful truncation of the DSEs.  That
entails many important results.\cite{mr97}

1) The axial-vector vertex has a pole at $P^2=-m_H^2$ whose residue is $f_H$,
the leptonic decay constant:
\begin{eqnarray}
\label{caint}
f_H P_\mu = 
Z_2\int^\Lambda_q\,\case{1}{2}
{\rm tr}\left[\left(T^H\right)^{\rm t} \gamma_5 \gamma_\mu 
{\cal S}(q_+) \Gamma_H(q;P) {\cal S}(q_-)\right]\,,
\end{eqnarray} 
with the trace over colour, Dirac and flavour indices.  

2) In the chiral limit
\begin{equation}
\label{bwti}
\begin{array}{ll}
f_H E_H(k;0)  =   B_0(k^2)\,,\; & 
F_R(k;0) +  2 \, f_H F_H(k;0)   =  A_0(k^2)\,, \\
G_R(k;0) +  2 \,f_H G_H(k;0)     =  2 A_0^\prime(k^2)\,,\; &
H_R(k;0) +  2 \,f_H H_H(k;0)     =  0\,,
\end{array}
\end{equation}
where $A_0(k^2)$ and $B_0(k^2)$ are the solutions of Eq.~(\ref{gendse}) in
the chiral limit, and $F_R$, $G_R$ and $H_R$ are calculable functions in
$\Gamma^H_{5\mu}$.  This shows that when chiral symmetry is dynamically
broken: 1) the flavour-nonsinglet, pseudoscalar BSE has a massless solution;
2) the Bethe-Salpeter amplitude for the massless bound state has a term
proportional to $\gamma_5$ alone, with the momentum-dependence of $E_H(k;0)$
completely determined by that of $B_0(k^2)$, in addition to terms
proportional to other pseudoscalar Dirac structures that are nonzero in
general; and 3) the axial-vector vertex, $\Gamma_{5 \mu}^H(k;P)$, is
dominated by the pseudoscalar bound state pole for $P^2\simeq 0$.  The
converse is also true.

3) The pseudoscalar vertex also has a pole at $P^2=-m_H^2$ whose residue is
\begin{equation}
\label{rH}
i r_H = Z_4\int^\Lambda_q\,\case{1}{2}
{\rm tr}\left[\left(T^H\right)^{\rm t} \gamma_5 
{\cal S}(q_+) \Gamma_H(q;P) {\cal S}(q_-)\right]\,.
\end{equation}

4) There is an identity between the residues of the pseudoscalar meson pole
in the axial-vector and pseudoscalar vertices that is satisfied independent
of the magnitude of the current quark mass:
\begin{equation}
\label{gmora}
f_H\,m_H^2 = r_H \, {\cal M}_H\,,\;\;
{\cal M}_H := {\rm tr}_{\rm flavour}
\left[M_{(\mu)}\,\left\{T^H,\left(T^H\right)^{\rm t}\right\}\right]\,.
\end{equation}
\subsection{Corollaries}\label{subsec:C1}
Equation~(\ref{gmora}) is a mass formula for flavour-octet pseudoscalar
mesons.  For small current-quark masses, using Eqs.~(\ref{genpibsa}) and
(\ref{bwti}), Eq.~({\ref{rH}) yields
\begin{equation}
\label{cbqbq}
\begin{array}{lcr}
\displaystyle
r_H^0  =  -\,\frac{1}{f_H^0}\, \langle \bar q q \rangle_\mu^0 \,
, & & 
\displaystyle
\,-\,\langle \bar q q \rangle_\mu^0 :=  
Z_4(\mu^2,\Lambda^2)\, N_c \int^\Lambda_q\,{\rm tr}_{\rm Dirac}
        \left[ S_{\hat m =0}(q) \right]\,,
\end{array}
\end{equation}
where $ \langle \bar q q \rangle_\mu^0 $ is the chiral limit {\it vacuum
quark condensate}, which is renormalisation-point dependent but independent
of the gauge parameter and the regularisation mass-scale:$\,$\cite{mr97}
$\langle \bar q q \rangle^0_{\mu=1\,{\rm GeV}} =
-\,(0.241\,{\rm GeV})^3\,.$
Now one obtains immediately from Eqs.~(\ref{gmora}) and (\ref{cbqbq})
\begin{eqnarray}
\label{gmorepi}
f_{\pi}^2 m_{\pi}^2 & = &-\,\left[m_u^\mu + m_d^\mu\right]
       \langle \bar q q \rangle_\mu^0 + {\rm O}\left(\hat m_q^2\right)\,\\
\label{gmoreKp}
f_{K^+}^2 m_{K^+}^2 & = &-\,\left[m_u^\mu + m_s^\mu\right]
       \langle \bar q q \rangle_\mu^0 + {\rm O}\left(\hat m_q^2\right)\,,
\end{eqnarray}
which exemplify what is commonly known as the Gell-Mann--Oakes--Renner
relation.  [$\hat m_q$ is the renormalisation-point-independent current-quark
mass.]  In a typical, model calculation$\,$\cite{mr97} Eq.~(\ref{gmorepi}) is
accurate to 7\%, while Eq.~(\ref{gmoreKp}) receives corrections of 43\% from
${\rm O}\left(\hat m_q^2\right)$ and higher, see Fig.~\ref{mHmq}.

In the heavy-quark limit, $\eta_P=1$ in Eq.~(\ref{genbse}) and the
heavy-meson velocity $(v_\mu)$ and binding energy $(E)$ are defined via: $P:=
m_H\,v_\mu$, $v^2=-1$, $M_H:= (\hat M_Q + E)$, with $M_H$ the heavy-meson
mass and $\hat M_Q \approx M^E_Q\approx \hat m_Q$; see Sec.~\ref{subsec:B2}.
In this case$\,$\cite{misha}, at leading order in $1/M_H$,
\begin{eqnarray}
\label{hQ}
S(q + P)  & = &  \case{1}{2}\,\frac{1 - i \gamma\cdot v}{q\cdot v - E}\,,\\
\label{hG}
\Gamma_H(q;P)  & = &  \sqrt{M_H}\,\hat\Gamma_H(q;P) \,,
\end{eqnarray}
where the canonical normalisation condition for $\hat\Gamma_H(q;P)$ is
independent of $M_H$.  Using Eqs.~(\ref{hQ}) and (\ref{hG}) in
Eq.~(\ref{caint}) one obtains
\begin{equation}
\label{fHhQ}
f_H \propto 1/\sqrt{ M_H}
% \frac{1}{\surd M_H}
\end{equation}
and this along with Eqs.~(\ref{rH}) and (\ref{gmora}) yields
\begin{equation}
\label{mHhQ}
M_H \propto \hat m_Q\,.
\end{equation}
A model study$\,$\cite{mr98} shows Eq.~(\ref{mHhQ}) to be valid for $\hat m_Q
\gsim \hat m_s$, and this is confirmed by data, Fig.~\ref{mHmq}.  
\begin{figure}[t]
\centering{\ \epsfig{figure=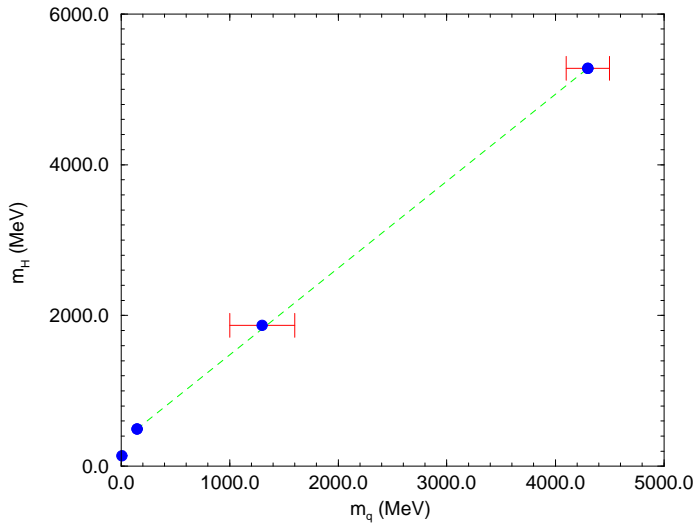,height=7.0cm} }\vspace*{-0.4\baselineskip}
\caption{Pseudoscalar meson mass as a function of the mass of the heaviest
constituent, $\hat m_q$.\protect\cite{pdg96} Only the $\pi$ does not lie on
the same straight line.
\label{mHmq}}
\end{figure}
However, both calculations and available data suggest that Eq.~(\ref{fHhQ})
is not manifest until $\hat m_Q > \hat m_c$.
\section{Finite $T$ and $\mu$}
The study of QCD at finite temperature and baryon number density proceeds via
the introduction of the intensive variables: temperature and quark chemical
potential.  These are additional mass-scales, with which the coupling can
{\it run} and hence, for $T\gg \Lambda_{\rm QCD}$ and/or $\mu\gg \Lambda_{\rm
QCD}$, $\alpha_{\rm S}(Q^2=0,T,\mu)\sim 0$.  It follows that at finite
temperature and/or baryon number density there is a phase of QCD in which
quarks and gluons are weakly interacting, {\em irrespective} of the momentum
transfer; i.e., a quark-gluon plasma.  Such a phase of matter existed
approximately one microsecond after the big-bang.  In this phase confinement
and DCSB are absent and the nature of the strong interaction spectrum is
qualitatively different.

Nonperturbative methods are necessary to study the phase transition, which is
characterised by qualitative changes in order parameters such as the quark
condensate.  One widely used approach is the numerical simulation of finite
temperature lattice-QCD, with the first simulations in the early eighties and
extensive efforts since then.\cite{karsch95} The commonly used quenched
approximation is inadequate for studying the phase diagram of finite
temperature QCD because the details of the transition depend sensitively on
the number of active (light) flavours.  It is therefore necessary to include
the fermion determinant.  

That is even more important in the presence of $\mu$, which modifies the
fermion piece of the Euclidean action: $\gamma\cdot \partial + m \to
\gamma\cdot \partial - \gamma_4 \mu + m$, and thus the fermion determinant
acquires an explicit imaginary part.  The $\mu\neq 0$ QCD action being
complex entails that the study of finite density is significantly more
difficult than that of finite temperature.  Simulations that ignore the
fermion determinant at $\mu\neq 0$ encounter a forbidden region, which begins
at $\mu = m_\pi/2$,\cite{dks96} and since $m_\pi\to 0$ in the chiral limit
this is a serious limitation, preventing a reliable study of chiral symmetry
restoration.  The phase of the fermion determinant is essential in
eliminating this artefact.\cite{adam}

The contemporary application of DSEs at finite temperature and chemical
potential is a straightforward extension of the $T=0=\mu$ studies.  The
direct approach is to develop a finite-$T$ extension of {\it Ans\"atze} for
the dressed-gluon propagator.  The quark DSE can then be solved and, having
the dressed-quark and -gluon propagators, the response of bound states to
increases in $T$ and $\mu$ can be studied.  As a nonperturbative approach
that allows the simultaneous study of DCSB and confinement, the DSEs have a
significant overlap with lattice simulations: each quantity that can be
estimated using lattice simulations can also be calculated using the DSEs.
This means they can be used to check the lattice simulations, and
importantly, that lattice simulations can be used to constrain their
model-dependent aspects.  Once agreement is obtained on the common domain,
the DSEs can be used to explore phenomena presently inaccessible to lattice
simulations.
\subsection{Finite-$(T,\mu)$ Quark DSE}
The renormalised dressed-quark propagator at finite-$(T,\mu)$ has the form
\begin{eqnarray}
\label{genformS}
S(\vec{p},\tilde\omega_k)  & = & \frac{1}
{i\vec{\gamma}\cdot \vec{p}\,A(\vec{p},\tilde\omega_k)
+i\gamma_4\,\tilde\omega_k C(\vec{p},\tilde\omega_k)+
B(\vec{p},\tilde\omega_k)}\\
& \equiv&  -i\vec{\gamma}\cdot \vec{p}\,\sigma_A(\vec{p},\tilde\omega_k)
-i\gamma_4\,\tilde\omega_k \sigma_C(\vec{p},\tilde\omega_k)+
\sigma_B(\vec{p},\tilde\omega_k)\, 
\end{eqnarray}
where $\tilde\omega_k := \omega_k + i \mu $ with $\omega_k= (2 k + 1)\,\pi
\,T$, the fermion Matsubara frequencies, $k\in {\rm Z}\!\!\!{\rm Z}$.  The
complex scalar functions: $A(\vec{p},\tilde\omega_k)$,
$B(\vec{p},\tilde\omega_k)$ and $C(\vec{p},\tilde\omega_k)$ satisfy:
$ {\cal F}(\vec{p},\tilde\omega_k)^\ast = {\cal
F}(\vec{p},\tilde\omega_{-k-1})\,, $
${\cal F}=A,B,C$, and although not explicitly indicated they are functions
only of $|\vec{p}|^2$ and $\tilde\omega_k^2$.  

$S(\vec{p},\tilde\omega_k)$ satisfies the DSE
\begin{equation}
\label{qDSE}
S^{-1}(\vec{p},\tilde\omega_k) = Z_2^A \,i\vec{\gamma}\cdot \vec{p} + Z_2 \,
(i\gamma_4\,\tilde\omega_k + m_{\rm bm})\, 
        + \Sigma^\prime(\vec{p},\tilde\omega_k)\,;
\end{equation}
where the regularised self energy is
\begin{eqnarray}
\Sigma^\prime(\vec{p},\tilde\omega_k) & = & i\vec{\gamma}\cdot \vec{p}
\,\Sigma_A^\prime(\vec{p},\tilde\omega_k) + i\gamma_4\,\tilde\omega_k
\,\Sigma_C^\prime(\vec{p},\tilde\omega_k) +
\Sigma_B^\prime(\vec{p},\tilde\omega_k)\; ,
\end{eqnarray}
\begin{equation}
\Sigma_{\cal F}^\prime(\vec{p},\tilde\omega_k)  = 
\int_{l,q}^{\bar\Lambda}\, \case{4}{3}\,g^2\,
D_{\mu\nu}(\vec{p}-\vec{q},\tilde\omega_k-\tilde\omega_l)\case{1}{4} {\rm
tr}\left[{\cal P}_{\cal F} \gamma_\mu
S(\vec{q},\tilde\omega_l)
\Gamma_\nu(\vec{q},\tilde\omega_l;\vec{p},\tilde\omega_k)\right]\,,
\label{regself}
\end{equation}
$\int_{l,q}^{\bar\Lambda}:=\, T
\,\sum_{l=-\infty}^\infty\,\int^{\bar\Lambda}\frac{d^3q}{(2\pi)^3}$ and
${\cal P}_A:= -(Z_1^A/p^2)i\gamma\cdot p$, ${\cal P}_B:= Z_1 $, ${\cal P}_C:=
-(Z_1/\tilde\omega_k)i\gamma_4$.

The finite-$(T,\mu)$, Landau-gauge dressed-gluon propagator has the form
\begin{eqnarray}
g^2 D_{\mu\nu}(\vec{p},\Omega) & = &
P_{\mu\nu}^L(\vec{p},\Omega) \,\Delta_F(\vec{p},\Omega) + 
P_{\mu\nu}^T(\vec{p})\, \Delta_G(p,\Omega) \,,\\
P_{\mu\nu}^T(\vec{p}) & := &\left\{
\begin{array}{ll}
0; \; & \mu\;{\rm and/or} \;\nu = 4,\\
\displaystyle
\delta_{ij} - \frac{p_i p_j}{|\vec{p}|^2}; \; & \mu,\nu=i,j=1,2,3
\end{array}\right.\,,
\end{eqnarray}
with $P_{\mu\nu}^T(p) + P_{\mu\nu}^L(p,p_4) = \delta_{\mu\nu}- p_\mu
p_\nu/{\sum_{\alpha=1}^4 \,p_\alpha p_\alpha}$; $\mu,\nu= 1,\ldots, 4$.  

In studying confinement one cannot assume that the analytic structure of a
dressed propagator is the same as that of the free particle propagator: it
must be determined dynamically.  The $\tilde
p_k:=(\vec{p},\tilde\omega_k)$-dependence of $A$ and $C$ is qualitatively
important since it can conspire with that of $B$ to eliminate free-particle
poles in the dressed-quark propagator.\cite{burden} In this case the
propagator does not have a Lehmann representation so that, in general, the
Matsubara sum cannot be evaluated analytically.  More importantly, it either
complicates or precludes a real-time formulation of the finite temperature
theory, which makes the study of nonequilibrium thermodynamics a very
challenging problem.  In addition, the $\tilde p_k$-dependence of $A$ and $C$
can be a crucial factor in determining the behaviour of bulk thermodynamic
quantities such as the pressure and entropy, being responsible for these
quantities reaching their respective ultrarelativistic limits only for very
large values of $T$ and $\mu$.  It is therefore important in any DSE study to
retain $A(\tilde p_k)$ and $C(\tilde p_k)$, and their dependence on $\tilde
p_k$.
\subsection{Phase Transitions and Order Parameters}
One order parameter for the chiral symmetry restoration transition is well
known - it is the quark condensate, defined via the renormalised
dressed-quark propagator, Eq.~(\ref{cbqbq}).  An equivalent order parameter
is
\begin{equation}
\label{chiorder}
{\cal X} := {\sf Re}\,B_0(\vec{p}=0,\tilde \omega_0)\,,
\end{equation}
which makes it clear that the zeroth Matsubara mode determines the character
of the chiral phase transition.  An order parameter for confinement, valid
for both light- and heavy-quarks, was introduced in Ref.~[\ref{prla}].  It is
a single, quantitative measure of whether or not a Schwinger function has a
Lehmann representation, and it has been used$\,$\cite{m95} to striking effect
in QED$_3$.
\subsection{Study at $(T\neq 0,\mu=0)$}\label{subsec:D1}
Deconfinement and chiral symmetry restoration have been studied$\,$\cite{prl}
in a DSE-model of two-light-flavour QCD.  The quark DSE was solved using a
one-parameter model$\,$\cite{fr96} dressed-gluon propagator, which provides a
good description of $\pi$ and $\rho$-meson observables at $T=0=\mu$.  The
transitions are coincident and second-order at a critical temperature of
$T_c\approx 150\,$MeV, with the same estimated critical exponents:
$\beta=0.33\pm 0.03$.  Both the pion mass, $m_\pi$, and the pion leptonic
decay constant, $f_\pi$, are insensitive to $T$ for $T\lsim 0.7\,T_c$.
However, as $T\to T_c$, the pion mass increases substantially, as thermal
fluctuations overwhelm quark-antiquark attraction in the pseudoscalar
channel, until, at $T=T_c$, $f_\pi\to 0$ and there is no bound state.  These
results confirm those of contemporary numerical simulations of finite-$T$
lattice-QCD~\cite{karsch95}.
\subsection{Complementary study at $(T= 0,\mu\neq 0)$}\label{subsec:D2}
The behaviour of this model at $\mu\neq 0$ has also been
explored.$\,$\cite{greg} The model employs a dressed-gluon propagator with
the simple form
\begin{eqnarray}
\label{gksquare}
\frac{{\cal G}(k^2)}{k^2} & = &
\case{16}{9} \pi^2 \left[ 4 \pi^2 m_t^2 \delta^4(k)
+ \frac{1- {\rm e}^{-[k^2/(4 m_t^2)]}}{k^2}\right]\,,
\end{eqnarray}
where$\,$\cite{fr96} $m_t=0.69\;$GeV is a mass-scale that marks the boundary
between the perturbative and nonperturbative domains, and the quark DSE was
solved in rainbow approximation.  The solution has the form
\begin{equation}
S(p_{[\mu]}):= -i
\vec{\gamma}\cdot \vec{p}\, \sigma_A(p_{[\mu]} ) 
        - i \gamma_4\, \omega_{[\mu]} \,
\sigma_C(p_{[\mu]}) + \sigma_B(p_{[\mu]})\,, 
\end{equation}
where $p_{[\mu]}:= (\vec{p},\omega_{[\mu]})$, with $\omega_{[\mu]} := p_4 + i
\mu$.  There are two distinct types of solution: a Nambu-Goldstone mode
characterised by $\sigma_{B_0} \not\equiv 0$; and a Wigner-Weyl mode
characterised by $\sigma_{B_0} \equiv 0$.

The possibility of a phase transition between the two modes is explored by
calculating the relative stability of the different phases, which is measured
by the difference between the tree-level auxiliary-field effective-action:
\begin{eqnarray}
\label{bagpres}
\lefteqn{\case{1}{2 N_f N_c}\,{\cal B}(\mu)  :=  }\\
&& 
\nonumber \int_p^\Lambda\,
\left\{ \ln\left[\frac{|\vec{p}|^2 A_0^2 + \omega_{[\mu]}^2 C_0^2 + B_0^2}
                {|\vec{p}|^2 \hat A_0^2 + \omega_{[\mu]}^2 \hat C_0^2}\right]
+  |\vec{p}|^2 \left(\sigma_{A_0} - \hat\sigma_{A_0}\right)
+  \omega_{[\mu]}^2 \left(\sigma_{C_0} - \hat\sigma_{C_0}\right)\right\}\,,
\end{eqnarray}
where $\hat A$ and $\hat C$ represent the solution of Eq.~(\ref{qDSE})
obtained when $B_0\equiv 0$; i.e., when DCSB is absent.  This solution exists
for all $\mu$.  ${\cal B}(\mu)$ defines a $\mu$-dependent ``bag
constant''.\cite{reg85} It is positive when the Nambu-Goldstone phase is
dynamically favoured; i.e., has the highest pressure, and becomes negative
when the Wigner pressure becomes larger.  Hence the critical chemical
potential is the zero of ${\cal B}(\mu)$, which is $\mu_c=375\,$MeV.  The
abrupt switch from the Nambu-Goldstone to the Wigner-Weyl mode signals a
first order transition.

The chiral order parameter {\it increases} with increasing chemical potential
up to $\mu_c $, with $\chi(\mu_c)/\chi(0)\approx 1.2$, whereas the
deconfinement order parameter, $\kappa(\mu)$, is insensitive to increasing
$\mu$.  At $\mu_c$ they both drop immediately and discontinuously to zero, as
expected of coincident, first-order phase transitions.  The increase of
${\cal X}$ with $\mu$ is a necessary consequence of the momentum dependence
of the scalar piece of the quark self energy, $B(p_{[\mu]})$.\cite{thermo} The
vacuum quark condensate behaves in qualitatively the same manner as $\chi$.

Even though the chiral order parameter {\it increases} with $\mu$, $m_\pi$
{\it decreases} slowly as $\mu$ increases, with $m_\pi(\mu\approx
0.7\,\mu_c)/m_\pi(0) \approx 0.94$.  At this point $m_\pi$ begins to increase
although, for $\mu<\mu_c$, $m_\pi(\mu)$ does not exceed $m_\pi(0)$, which
precludes pion condensation.  The behaviour of $m_\pi$ results from mutually
compensating increases in $f_\pi^2$ and $\langle m_R^\zeta (\bar q
q)_\zeta\rangle_\pi$.  $f_\pi$ is insensitive to the chemical potential until
$\mu\approx 0.7\,\mu_c$, when it increases sharply so that
$f_\pi(\mu_c^-)/f_\pi(\mu=0) \approx 1.25$.  The relative insensitivity of
$m_\pi$ and $f_\pi$ to changes in $\mu$, until very near $\mu_c$, mirrors the
behaviour of these observables at finite-$T$.\cite{prl} For example, it leads
only to a $14$\% increase in the $\pi\to \mu\nu$ decay width at $\mu\approx
0.9\,\mu_c$.  The universal scaling conjecture of Ref.~[\ref{brownR}] is
inconsistent with the anticorrelation observed between the $\mu$-dependence
of $f_\pi$ and $m_\pi$.

Comparing the $\mu$-dependence of $f_\pi$ and $m_\pi$ with their
$T$-dependence, one observes an anticorrelation; e.g., at $\mu=0$, $f_\pi$
falls continuously to zero as $T$ is increased towards $T_c \approx
150\,$MeV.  This is a necessary consequence of the momentum-dependence of the
quark self energy.\vspace*{-0.2\baselineskip}
\addtocounter{enumi}{1}
\begin{description}
\item[Note~\Alph{enumi}:] In calculating these observables one obtains
expressions for $m_\pi^2$ or $f_\pi^2$ and hence the natural dimension is
mass-squared.  Therefore their behaviour at finite $T$ and $\mu$ is
determined by
\begin{equation}
{\sf Re}(\omega_{[\mu]}^2) \sim [\pi^2 T^2 - \mu^2]\,,
\end{equation}
where the $T$-dependence arises from the introduction of the fermion
Matsubara frequency: \mbox{$p_4 \to (2k +1)\pi T$}.  Hence when such a
quantity decreases with $T$ it will increase with $\mu$, and
vice-versa.\cite{schmidt98} \vspace*{-0.2\baselineskip}
\end{description}

The confined-quark vacuum consists of quark-antiquark pairs correlated in a
scalar condensate.\vspace*{-0.2\baselineskip}
\addtocounter{enumi}{1}
\begin{description}
\item[Note~\Alph{enumi}:] Increasing $\mu$ increases the scalar
density: $(-\langle \bar q q \rangle)$.  This result is an expected
consequence of confinement, which entails that each additional quark must be
locally paired with an antiquark thereby increasing the density of condensate
pairs as $\mu$ is increased.\vspace*{-0.2\baselineskip}
\end{description}
For this reason, as long as $\mu<\mu_c$, there is no excess of particles over
antiparticles in the vacuum and hence the baryon number density remains
zero;\cite{thermo} i.e., $ \rho_B^{u+d}=0\,,\;  \mu < \mu_c $.  This
is just the statement that quark-antiquark pairs confined in the condensate
do not contribute to the baryon number density.

The quark pressure, $P^{u+d}[\mu]$, can be calculated$\,$\cite{thermo} and
one finds that after deconfinement it increases rapidly, as the condensate
``breaks-up'', and an excess of quarks over antiquarks develops.  The
baryon-number density, \mbox{$\rho_B^{u+d} = (1/3)\partial P^{u+d}/\partial
\mu$}, also increases rapidly, with
\begin{equation}
\rho_B^{u+d}(\mu\approx 2 \mu_c) \simeq 3 \,\rho_0\,,
\end{equation}
where $\rho_0=0.16\,{\rm fm}^{-3}$ is the equilibrium density of nuclear
matter.  For comparison, the central core density expected in a
$1.4\,M_\odot$ neutron star is $3.6$-$4.1\,\rho_0$.\cite{wiringa}
Finally,
at $\mu\sim 5 \mu_c$, the quark pressure saturates the ultrarelativistic
limit: $P^{u+d}= \mu^4/(2\pi^2)$, and there is a simple relation between
baryon-density and chemical-potential:
\begin{equation}
\label{fqg}
\rho_B^{u_F+d_F}(\mu) = \frac{1}{3} \, \frac{2 \mu^3}{\pi^2}\,,
\; \forall \mu \gsim 5 \mu_c \,,
\end{equation}
so that $\rho_B^{u_F+d_F}(5\mu_c)\sim 350\,\rho_0$.  Thus the quark pressure
in the deconfined domain overwhelms any finite, additive contribution of
hadrons to the equation of state.  That was anticipated in Ref.~[\ref{gregR}]
where the hadron contribution was neglected.  This discussion suggests that a
quark-gluon plasma may be present in the core of dense neutron stars.
\subsection{Simultaneous study of $(T\neq 0,\mu\neq 0)$}
This is a difficult problem and the most complete study$\,$\cite{thermo} to
date employs a simple {\it Ansatz} for the dressed-gluon propagator that
exhibits the infrared enhancement suggested by Ref.~[\ref{bp89R}]:
\begin{equation}
\label{mnprop}
g^2 D_{\mu\nu}(\vec{p},\Omega_k) = 
\left(\delta_{\mu\nu} 
- \frac{p_\mu p_\nu}{|\vec{p}|^2+ \Omega_k^2} \right)
2 \pi^3 \,\frac{\eta^2}{T}\, \delta_{k0}\, \delta^3(\vec{p})\,,
\end{equation}
with $\Omega_k=2 k \pi T$, the boson Matsubara frequency.  As an
infrared-dominant model that does not represent well the behaviour of
$D_{\mu\nu}(\vec{p},\Omega_k)$ away from $|\vec{p}|^2+ \Omega_k^2 \approx 0$,
some model-dependent artefacts arise.  However, there is significant merit in
its simplicity and, since the artefacts are easily identified, the model
remains useful as a means of elucidating many of the qualitative features of
more sophisticated {\it Ans\"atze}.

With this model, using the rainbow approximation, the quark DSE
is$\,$\cite{bender96}
\begin{equation}
\label{mndse}
S^{-1}(\vec{p},\omega_k) = S_0^{-1}(\vec{p},\tilde \omega_k)
        + \case{1}{4}\eta^2\gamma_\nu S(\vec{p},\tilde \omega_k) \gamma_\nu\,.
\end{equation}
A simplicity inherent in Eq.~(\ref{mnprop}) is now apparent: it allows the
reduction of an integral equation to an algebraic equation, in whose solution
many of the qualitative features of more sophisticated models are manifest.

In the chiral limit Eq.~(\ref{mndse}) reduces to a quadratic equation for
$B(\tilde p_k)$, which has two qualitatively distinct solutions.  The
Nambu-Goldstone solution, with
\begin{eqnarray}
\label{ngsoln}
B(\tilde p_k) & = &\left\{
\begin{array}{lcl}
\sqrt{\eta^2 - 4 \tilde p_k^2}\,, & &{\sf Re}(\tilde p_k^2)<\case{\eta^2}{4}\\
0\,, & & {\rm otherwise}
\end{array}\right.\\
C(\tilde p_k) & = &\left\{
\begin{array}{lcl}
2\,, & & {\sf Re}(\tilde p_k^2)<\case{\eta^2}{4}\\
\case{1}{2}\left( 1 + \sqrt{1 + \case{2 \eta^2}{\tilde p_k^2}}\right)
\,,& & {\rm otherwise}\,,
\end{array}\right.
\end{eqnarray}
describes a phase of this model in which: 1) chiral symmetry is dynamically
broken, because one has a nonzero quark mass-function, $B(\tilde p_k)$, in
the absence of a current-quark mass; and 2) the dressed-quarks are confined,
because the propagator described by these functions does not have a Lehmann
representation.  The alternative Wigner solution, for which
\begin{eqnarray}
\label{wsoln}
\hat B(\tilde p_k)  \equiv  0 &,\;& 
\hat C(\tilde p_k)  = 
\case{1}{2}\left( 1 + \sqrt{1 + \case{2 \eta^2}{\tilde p_k^2}}\right)\,,
\end{eqnarray}
describes a phase of the model with neither DCSB nor confinement.
 
The relative stability of the different phases is measured by a
$(T,\mu)$-dependent vacuum pressure difference, which in the chiral limit is
\begin{eqnarray}
\label{bagcons}
\lefteqn{{\cal B}(T,\mu)  = }\\
& & \nonumber
 \eta^4\,2 N_c N_f \frac{\bar T}{\pi^2}\sum_{l=0}^{l_{\rm max}}
        \int_0^{\bar\Lambda_l}\,dy\,y^2\,
        \left\{{\sf Re}\left(2 \bar p_l^2 \right) 
                - {\sf Re}\left(\frac{1}{C(\bar p_l)}\right)
- \ln\left| \bar p_l^2 C(\bar p_l)^2\right|        \right\},
\end{eqnarray}
with: $\bar T=T/\eta$, $\bar \mu=\mu/\eta$; $l_{max}$ is the largest value of
$l$ for which $\bar\omega^2_{l_{\rm max}}\leq (1/4)+\bar\mu^2$ and this
also specifies $\omega_{l_{max}}$, $\bar\Lambda^2 = \bar\omega^2_{l_{\rm
max}}-\bar\omega_l^2$, $\bar p_l = (\vec{y},\bar\omega_l+i\bar\mu)$.  The
condition ${\cal B}(T,\mu) \equiv 0$ defines the phase boundary in the
$(\mu,T)$-plane.

Again, the deconfinement and chiral symmetry restoration transitions are
coincident.  For $\mu=0$ the transition is second order and the critical
temperature is $T_c^0 = 0.159\,\eta$, which using the value of
$\eta=1.06\,$GeV obtained by fitting the $\pi$ and $\rho$ masses corresponds
to $T_c^0 = 0.170\,$GeV.  This is only 12\% larger than the value reported in
Sec.~\ref{subsec:D1}, and the order of the transition is the same.  For any
$\mu \neq 0$ the transition is first-order.  For $T=0$ the critical chemical
potential is $\mu_c^0=0.3\,$GeV, which is \mbox{$\approx 30$\%} smaller than
the result in Sec.~\ref{subsec:D2}.  The discontinuity in the order
parameters vanishes as $\mu\to 0$.

The quark pressure, $P_q$, is calculated easily in this model.  Confinement
means that $P_q \equiv 0$ in the confined domain.  In the deconfined domain
it approaches
\begin{equation}
P_{\rm UR}:= \label{sbpres}
        \eta^4\, N_c N_f \frac{1}{12\pi^2}\left(
        \bar\mu^4 + 2 \pi^2 \bar\mu^2 \bar T^2 + \frac{7}{15}\pi^4
                        \bar T^4\right)\,, 
\end{equation}
the ultrarelativistic, free particle limit, at large values of $\bar T$ and
$\bar\mu$.  The approach to this limit is slow, however.  For example, at
$\bar T \sim 0.3 \sim 2 \bar T_c^0$, or $\bar \mu \sim 1.0 \sim 3
\bar\mu_c^0$, $P_q$ is only $0.5\,P_{\rm UR}$.  A qualitatively similar
result is observed in numerical simulations of finite-$T$
lattice-QCD.\cite{karsch95} This feature results from the persistence of the
momentum dependent modifications of the quark propagator into the deconfined
domain, and predicts that there is a ``mirroring'' of finite-$T$ behaviour in
the dependence of the bulk thermodynamic quantities on $\mu$.
\subsection{$\pi$ and $\rho$ properties}
The model discussed in the last section has been used$\,$\cite{schmidt98} to
study the $(T,\mu)$-dependence of $\pi$ and $\rho$ properties, and to
elucidate other features of models that employ a more sophisticated {\it
Ansatz} for the dressed-gluon propagator.  For example, the vacuum quark
condensate takes the simple form
\begin{equation}
\label{qbq}
-\langle \bar q q \rangle = 
\eta^3\,\frac{8 N_c}{\pi^2} \bar T\,\sum_{l=0}^{l_{\rm max}}\,
\int_0^{\bar\Lambda_l}\,dy\, y^2\,
{\sf Re}\left( \sqrt{\case{1}{4}- y^2 - \tilde\omega_{l}^2 }\right)\,:
\end{equation}
at $T=0=\mu$, $(-\langle \bar q q \rangle) = \eta^3 /(80\,\pi^2) = (0.11\,
\eta)^3$.  $(-\langle \bar q q \rangle)$ decreases with $T$ but {\it
increases} with $\mu$, up to a critical value of $\mu_c(T)$ when it drops
discontinuously to zero, in agreement with the behaviour reported in
Secs.~\ref{subsec:D1} and \ref{subsec:D2}, see Note~B.  This
vacuum rearrangement is manifest in the behaviour of the
necessarily-momentum-dependent scalar part of the quark self energy,
$B(\tilde p_k)$.

The leptonic decay constant also has a simple form in the chiral limit:
\begin{eqnarray}
\label{npialg}
f_\pi^2 & = & \eta^2 \frac{16 N_c }{\pi^2} \bar T\,\sum_{l=0}^{l_{\rm max}}\,
\frac{\bar\Lambda_l^3}{3} \left( 1 + 4 \,\bar\mu^2 - 4 \,\bar\omega_l^2 -
\case{8}{5}\,\bar\Lambda_l^2 \right)\,.
\end{eqnarray}
Characteristic in Eqs.~(\ref{qbq}) and (\ref{npialg}) is the combination
$\mu^2 - \omega_l^2$, see Note~A.  Without calculation, Eq.~(\ref{npialg})
indicates that $f_\pi$ will {\it decrease} with $T$ and {\it increase} with
$\mu$.  This provides a simple elucidation of the results described in
Secs.~\ref{subsec:D1} and \ref{subsec:D2}

The $(T,\mu)$-response of the $\pi$ and $\rho$ masses is determined by the
BSE
\begin{equation}
\label{bse}
\Gamma_M(\tilde p_k;\check P_\ell)= - \frac{\eta^2}{4}\,
{\sf Re}\left\{\gamma_\mu\,
S(\tilde p_i +\case{1}{2} \check P_\ell)\,
\Gamma_M(\tilde p_i;\check P_\ell)\,
S(\tilde p_i -\case{1}{2} \check P_\ell)\,\gamma_\mu\right\}\,,
\end{equation}
where $\check P_\ell := (\vec{P},\Omega_\ell)$, with the bound state mass
obtained by considering $\check P_{\ell=0}$.

The $\pi$ equation admits the solution
\begin{equation} 
\Gamma_\pi(P_0) = \gamma_5 \left(i \theta_1 
        + \vec{\gamma}\cdot \vec{P} \,\theta_2 \right)
\end{equation}
and the calculated $(T,\mu)$-dependence of the mass is depicted in
Fig.~\ref{pirhomass}.
\begin{figure}
\centering{\
\epsfig{figure=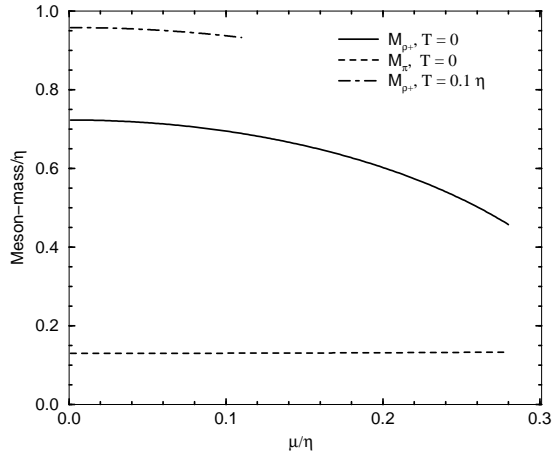,height=7.0cm}}\vspace*{-1.5\baselineskip}
\caption{\label{pirhomass} $M_{\rho+}$ and $m_\pi$ as a function of $\bar\mu$
for $\bar T = 0, 0.1$.  On the scale of this figure, $m_\pi$ is insensitive
to this variation of $T$.  The current-quark mass is $m= 0.011\,\eta$, which
for $\eta=1.06\,$GeV yields $M_{\rho+}= 770\,$MeV and $m_\pi=140\,$MeV at
$T=0=\mu$.}
\end{figure}

For the $\rho$-meson there are two components: one longitudinal and one
transverse to $\vec{P}$.  The solution of the BSE has the form
\begin{equation}
\Gamma_\rho = \left\{
\begin{array}{l}
\gamma_4 \,\theta_{\rho+} \\
\left(
\vec{\gamma} - \case{1}{|\vec{P}|^2}\,\vec{P} \vec{\gamma}\cdot\vec{P}\right)\,
        \theta_{\rho-}
\end{array}
\right.\,,
\end{equation}
where $\theta_{\rho+}$ labels the longitudinal and $\theta_{\rho-}$ the
transverse solution.  The eigenvalue equation obtained from (\ref{bse}) for
the bound state mass, $M_{\rho\pm}$, is
\begin{equation}
\label{rhomass}
\frac{\eta^2}{2}\,{\sf Re}\left\{ \sigma_S(\omega_{0+}^2
        - \case{1}{4} M_{\rho\pm}^2)^2 
- \left[ \pm \,\omega_{0+}^2 - \case{1}{4} M_{\rho\pm}^2\right]
        \sigma_V(\omega_{0+}^2- \case{1}{4} M_{\rho\pm}^2)^2 \right\}
= 1\,.
\end{equation}

The equation for the transverse component is obtained with $[- \omega_{0+}^2
- (1/4) M_{\rho-}^2]$ in (\ref{rhomass}).  Using the chiral-limit solutions,
Eq.~(\ref{ngsoln}), one obtains immediately that
\begin{equation}
M_{\rho-}^2 = \case{1}{2}\,\eta^2,\;\mbox{{\it independent} of $T$ and $\mu$.}
\end{equation}
Even for nonzero current-quark mass, $M_{\rho-}$ changes by less than 1\% as
$T$ and $\mu$ are increased from zero toward their critical values.  Its
insensitivity is consistent with the absence of a constant mass-shift in the
transverse polarisation tensor for a gauge-boson.

For the longitudinal component one obtains in the chiral limit:
\begin{equation}
\label{mplus}
M_{\rho+}^2 = \case{1}{2} \eta^2 - 4 (\mu^2 - \pi^2 T^2)\,.
\end{equation}
The characteristic combination $[\mu^2 - \pi^2 T^2]$ again indicates the
anticorrelation between the response of $M_{\rho+}$ to $T$ and its response
to $\mu$, and, like a gauge-boson Debye mass, that $M_{\rho+}^2$ rises
linearly with $T^2$ for $\mu=0$.  The $m\neq 0$ solution of
Eq.~(\ref{rhomass}) for the longitudinal component is plotted in
Fig.~\ref{pirhomass}: $M_{\rho+}$ {\it increases} with increasing $T$ and
{\it decreases} as $\mu$ increases.

Equation~(\ref{rhomass}) can also be applied to the $\phi$-meson.  The
transverse component is insensitive to $T$ and $\mu$, and the behaviour of
the longitudinal mass, $M_{\phi+}$, is qualitatively the same as that of the
$\rho$-meson: it increases with $T$ and decreases with $\mu$.  Using $\eta =
1.06\,$GeV, the model yields $M_{\phi\pm} = 1.02\,$GeV for $m_s = 180\,$MeV
at $T=0=\mu$.

In a 2-flavour, free-quark gas at $T=0$ the baryon number density is $\rho_B=
2 \mu^3/(3 \pi^2)\,$, by which gauge nuclear matter density,
$\rho_0=0.16\,$fm$^{-3}$, corresponds to $\mu= \mu_0 := 260\,$MeV$\,=
0.245\,\eta$.  At this chemical potential the algebraic model yields
\begin{eqnarray}
\label{mrhoa}
M_{\rho+}(\mu_0)  \approx  0.75 M_{\rho+}(\mu=0) &,\; &
M_{\phi+}(\mu_0)  \approx  0.85 M_{\phi+}(\mu=0)\,.
\end{eqnarray}
The study summarised in Sec.~\ref{subsec:D2} indicates that a better
representation of the ultraviolet behaviour of the dressed-gluon propagator
expands the horizontal scale in Fig.~\ref{pirhomass}, with the critical
chemical potential increased by 25\%.  This suggests that a more realistic
estimate is obtained by evaluating the mass at $\mu_0^\prime=0.20\,\eta$,
which yields
\begin{eqnarray}
\label{mrhob}
M_{\rho+}(\mu_0^\prime) \approx  0.85 M_{\rho+}(\mu=0) &,\; &
M_{\phi+}(\mu_0^\prime) \approx  0.90 M_{\phi+}(\mu=0) \,;
\end{eqnarray}
a small, quantitative modification.  The difference between
Eqs.~(\ref{mrhoa}) and (\ref{mrhob}) is a measure of the theoretical
uncertainty in the estimates in each case.  At the critical chemical
potential for $T=0$, $M_{\rho+} \approx 0.65\, M_{\rho+}(\mu=0)$ and
$M_{\phi+} \approx 0.80\, M_{\phi+}(\mu=0)$.

This simple model preserves the momentum-dependence of gluon and quark
dressing, which is an important qualitative feature of more sophisticated
studies.  Its simplicity means that many of the consequences of that dressing
can be demonstrated algebraically.  For example, it elucidates the origin of
an anticorrelation, found for a range of quantities, between their response
to increasing $T$ and that to increasing $\mu$.  And the $(T,\mu)$-dependence
of $(-\langle \bar q q)\rangle$ and $f_\pi$, understood algebraically, is
opposite to that observed for $m_{\rho+}$, hence the scaling law conjectured
in Ref.~[\ref{brownR}] is inconsistent with this calculation, as it is with
others of this type.
\section{Concluding Remarks}
This contribution illustrates the contemporary application of Dyson-Schwinger
equations to the analysis of observable strong interaction phenomena,
highlighting positive aspects and successes.  Many recent, interesting
studies have been neglected: calculations of the cross section for
diffractive, vector meson electroproduction,\cite{pichowsky} the electric
dipole moment of the $\rho$-meson,\cite{martin} and the electromagnetic pion
form factor;$\,$\cite{mrpion} an exploration of $\eta$-$\eta^\prime$
mixing;$\,$\cite{klabucar} and others reviewed in Ref.~[\ref{pctrevR}].
However, a simple enquiry of
\begin{center}
``http://xxx.lanl.gov/find/hep-ph'' 
\end{center}
with the keywords: ``Dyson-Schwinger'' or ``Schwinger-Dyson'', will provide a
guide to other current research.

In all phenomenological applications, modelling is involved, in particular of
the behaviour of the dressed Schwinger functions in the infrared.  (The
ultraviolet behaviour is fixed because of the connection with perturbation
theory.)  This is tied to the need to make truncations in order to define a
tractable problem.  Questions will always be asked regarding the fidelity of
this modelling.  The answers can only come slowly as, for example, more is
learnt about the constraints that Ward Identities and Slavnov-Taylor
identities in the theory can provide.  That approach has been particularly
fruitful in QED,\cite{ayse97} and already in the development of a systematic
truncation procedure for the kernel of the quark DSE and meson
BSE.\cite{bender96} In the meantime, and as is common, phenomenological
applications provide a key to understanding which elements of the approach
need improvement: the approach itself must also be explored under extreme
conditions.

\section*{Acknowledgments}
We are grateful to the staff of the Special Research Centre for the Subatomic
Structure of Matter at the University of Adelaide for their hospitality and
support during this workshop.  This work was supported by the US Department
of Energy, Nuclear Physics Division, under contract no. W-31-109-ENG-38.

\begin{flushleft}

\end{flushleft}

\end{document}